\documentclass[review,number,sort&compress]{elsarticle}

\usepackage{amssymb}

\usepackage{lineno}
\usepackage{footnote}
\usepackage{hyperref}

\journal{Nuclear Physics A}

\begin{document}

\begin{frontmatter}



\title{Development of Edgeless n-on-p Planar Pixel Sensors for future ATLAS Upgrades}


\author[1]{M.~BOMBEN\corref{cor1}}
\author[2]{A.~BAGOLINI}
\author[2]{M.~BOSCARDIN}
\author[3]{L.~BOSISIO}
\author[1,4]{G.~CALDERINI}
\author[1]{J.~CHAUVEAU}
\author[2]{G.~GIACOMINI}
\author[5]{A.~LA ROSA}
\author[1]{G.~MARCHIORI}
\author[2]{N.~ZORZI}

\address[1]{Laboratoire de Physique Nucleaire et de Hautes \'Energies (LPNHE)\\  Paris, France\\}
\address[2]{Fondazione Bruno Kessler, Centro per i Materiali e i Microsistemi (FBK-CMM)\\   Povo di Trento (TN), Italy}
\address[3]{Universit\`a di Trieste, Dipartimento di Fisica and INFN, Trieste, Italy}
\address[4]{Dipartimento di Fisica E. Fermi, Universit\`a di Pisa, and INFN Sez. di Pisa, Pisa, Italy}
\address[5]{Section de Physique (DPNC), Universit\'e de Gen\`eve, Gen\`eve, Switzerland}

\cortext[cor1]{corresponding author}

\begin{abstract}
The development of n-on-p ``edgeless'' planar pixel sensors being fabricated at FBK (Trento, Italy), 
 aimed at the upgrade of the ATLAS Inner Detector for the High Luminosity phase of the Large Hadron Collider (HL-LHC), is reported.
A characterizing feature of the devices is the reduced dead area at the edge, achieved by adopting the ``active edge'' technology, based on a deep etched trench, 
suitably doped to make an ohmic contact to the substrate.
The project is presented, along with the active edge process, the sensor design for this first n-on-p production and a selection of simulation results, including the expected 
charge collection efficiency after radiation fluence of $1 \times 10^{15} {\rm n_{ eq}}/{\rm cm}^2$ 
comparable to those expected at HL-LHC (about ten years of running, with an  integrated luminosity 
of 3000 fb$^{-1}$) for the outer pixel layers. We show that, after irradiation and at a bias voltage of 500~V, more than 50~\% of the  signal should be collected 
in the edge region; this 
confirms the validity of the active edge approach.
 \end{abstract}


\end{frontmatter}

\linenumbers


\section{Introduction}
\label{sec:intro}

Planar pixel sensors are nowadays the standard choice for particle tracking and vertex reconstruction in high energy physics experiments. 
The ATLAS collaboration will upgrade its current Pixel Detector~\cite{pixel-electronics-paper} in two phases: in 2013-2014 an additional $4^{\rm th}$ pixel layer 
will be inserted 
(Insertable B-Layer, IBL)~\cite{IBL}, while for the High Luminosity phase of LHC (HL-LHC)~\cite{HL-LHC} (beyond 2020) a completely new Pixel Detector is envisaged. 
The new pixel sensors will integrate a fluence of about $10^{16} {\rm n_{eq}}/{\rm cm}^2$ in the innermost layer, 
down to a fluence of $\phi =  1 \times 10^{15} \rm{n_{eq}/cm^2}$ in the mid-outer layers, 
for an integrated luminosity of 3000 fb$^{-1}$,  
with an instantaneous luminosity 
of $10^{35} {\rm cm}^{-2}{\rm s}^{-1}$. These harsh conditions demand radiation-hard devices and a finely segmented detector to cope with the expected 
high occupancy.
Several options are under investigation for the upgrade of the ATLAS pixel detector, including diamond~\cite{bib:Diamond}, silicon 3D~\cite{bib:3D}, HV-CMOS~\cite{bib:HVCMOS}
 and planar sensors~\cite{bib:PPS}. 

The new pixel sensors will need to have  
  high geometrical acceptance: the future material budget restrictions and  
 tight mechanical constraints require a geometry inefficiency below 2.5\%~\cite{IBL}.
 For example, the inactive areas at the device periphery should be less than 450~$\mu$m wide for IBL sensors~\cite{IBL}. 
In conventional sensor designs there is a relatively large un-instrumented area at the edge of the sensor to prevent the electric field from reaching the rim, where a large
 number of defects are present due to the wafer cutting; 
 for example the current ATLAS pixel sensor has an un-instrumented region of 1.1~mm at the edge~\cite{pixel-electronics-paper}, including Guard Rings (GRs).  GRs, placed all 
 around the pixel area, can help to improve the voltage-handling capability.

One way to reduce or even eliminate the insensitive region along the device periphery is offered by
 the ``active edge'' technique~\cite{bib:Kenney}, in which a deep vertical trench is etched along the device periphery throughout the entire wafer thickness, 
thus performing a damage free cut (this requires using a support wafer, to prevent the individual chips from getting loose). 
The trench is  then heavily doped, extending the ohmic back-contact to the lateral sides of the device: the depletion region can then extend to the edge without causing 
a large current increase.
 This is the technology we have chosen for realizing n-on-p pixel sensors with reduced inactive zone. 

Since a high bias voltage is required after heavy irradiation to maintain a deep depletion region and to ensure efficient charge collection in the presence of trapping effects, 
several GRs are commonly used. However, adding one or more GRs spoils the gain of the active behavior of the sensor edge. Indeed, once carriers are created
 in the guard ring area, 
or even outside,  electrons are collected by the guard rings, which, since are floating, re-emit the 
charge toward all the pixels next to them.
Such  charge, being distributed among a large  number of pixels, beyond the fact that it is insufficient to trigger a hit, cannot give any information about 
the hit position.
A compromise should be found between minimization of pixel to trench distance and presence of one or more GRs. In order to gain a  better insight into this point, 
detectors with different termination structures, with and without GRs,  have been simulated, designed and fabricated.

In this paper the active edge technology (Section~\ref{sec:active}) is presented; it has been chosen for a first production of  n-on-p planar sensors at FBK  (Section~\ref{sec:prod}). 
Studies performed with TCAD simulation tools (Section~\ref{sec:simu}) helped in defining the layout  and making a first estimation of the charge collection efficiency 
expected after irradiation.

\section{The active edge sensor fabrication at FBK}
\label{sec:active}

The sensors are fabricated on 100~mm diameter, high resistivity, p-type, Float Zone (FZ), \textless100\textgreater\, oriented, 200~${\rm \mu}$m thick wafers. 
The active edge technology~\cite{bib:Kenney} is used, which is a single sided process, featuring a doped trench, 
extending all the way through the wafer thickness, 
and completely surrounding the sensor. For mechanical reasons, a support wafer is therefore needed, making the back inaccessible after wafer-bonding.
Thus, as first process steps, a uniform high-dose boron implant has been performed on the back side, followed by a thermal oxide growth on 
both sides.

The wafers have then been shipped to Sintef~\cite{bib:sintef}, to be wafer-bonded to a 500~${\rm \mu}$m thick silicon substrate.
After having received back the wafers, the remaining process has been performed in the FBK clean-room.
A solution will be identified  to remove the support wafer and separate the devices\footnote{The easiest solution is the lapping of the support wafer once the sensors are diced.}; 
for the time being  all the detectors have been designed in order to allow conventional 
 saw-cut separation and substrate biasing from the front-side, through a dedicated ohmic contact (``bias tab'') . In this way, the efficiency  of the edge region before and after 
 irradiation can be studied even 
 without removing 
  the support wafer.

Up to the trench definition, the process follows standard steps. Since the read-out electrodes are n-type, they will be shorted together by the electron 
inversion layer, induced by the positive fixed charge present in the oxide, unless a p-type implant, compensating such charge, surrounds the pixels.
Both homogeneous (``p-spray'') and patterned (``p-stop'') implants have been used; 
the process splittings adopted in the fabrication batch only concern the presence and the doses of these implants,  as detailed in table~\ref{tab:isolation}. 
Electrical tests on irradiated devices will tell which combination can better guarantee both junction isolation and high breakdown voltages (which are competing demands, 
since they are in favor of
high and low boron doses, respectively), even after years of operation in a harsh radiation environment.

\begin{table}[!ht]
\begin{center}
\begin{tabular}{cc}
p-spray    & p-stop \\
\hline
	low dose &      absent\\
	high	 dose  &	 absent\\
	low   dose &     present\\
	high  dose  & 	present\\
	absent	& 	present
\end{tabular}
\end{center}
\caption{\label{tab:isolation}List of the different isolation solutions adopted in the process.}
\end{table}

Two patterned high dose implants, a  phosphorus implant forming the pixel and GR junctions  and a boron implant for  the ohmic contact
 to the substrate (``bias tab''), are then performed. 

The etching of the trench is accomplished by a Deep Reactive Ion Etching (DRIE) machine, the same used for the fabrication of 
3D detectors~\cite{bib:3DFBK}.
In the latter case, a 10~${\rm \mu}$m diameter and 200~${\rm \mu}$m deep hole has to be etched; the etching mask is made  by multiple stacks of dielectrics
 (oxide and nitride) plus a thick  photoresist. 
The trenches in an active edge sensor must be fully passing, {\it i.e.} their bottom  has to reach  the silicon oxide, which separates the active wafer from the 
support wafer.
In the etching of a trench, a problem arises from the fact that the photoresist tends to wear out and/or lift along the sides  of the trench, and then to be
  a less effective mask. Thicker stacks of dielectrics are thus needed.

\begin{figure}[!ht]
\begin{center}
\includegraphics[width=0.40\textwidth]{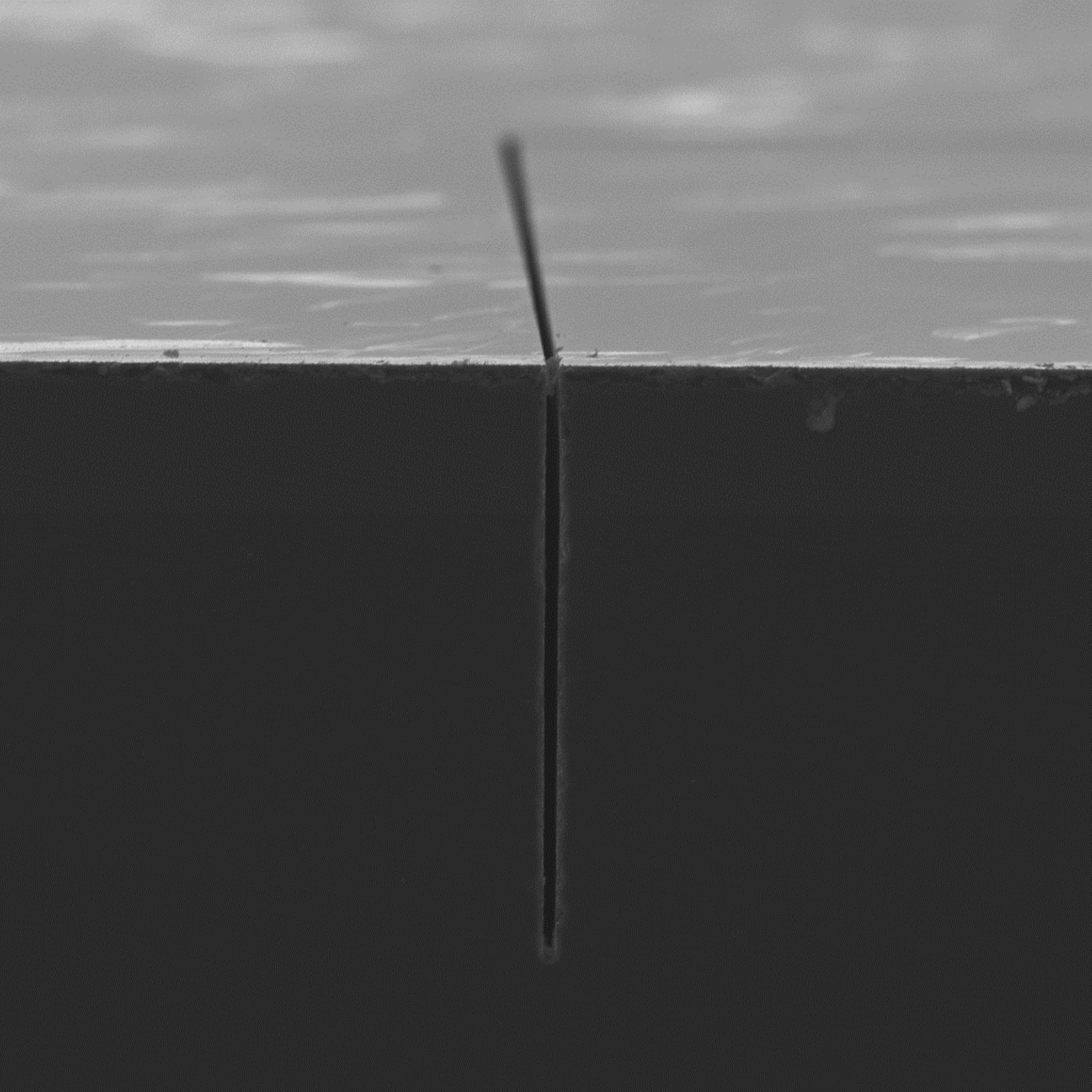}
\caption{\label{fig:trench}SEM picture of  a test trench, after cleaving the wafer perpendicularly to the surface and to the trench itself.}
\end{center}
\end{figure}

After the trench is etched, its walls are boron-doped in a diffusion furnace. 
Thus, a continuous ohmic contact to the substrate  is created on the trench wall and to the backside.
 FBK technology can routinely obtain very uniform, well defined and narrow trenches, as shown for example in Figure~\ref{fig:trench}. For a 200~${\rm \mu}$m thick bulk 
 the typical trench width is of~5~${\rm \mu}$m.
 Functional planar p-on-n devices with active edge have been already fabricated and reported~\cite{bib:3FBKtrenchponn}.

\begin{figure}[!ht]
\begin{center}
\includegraphics[width=0.89\textwidth]{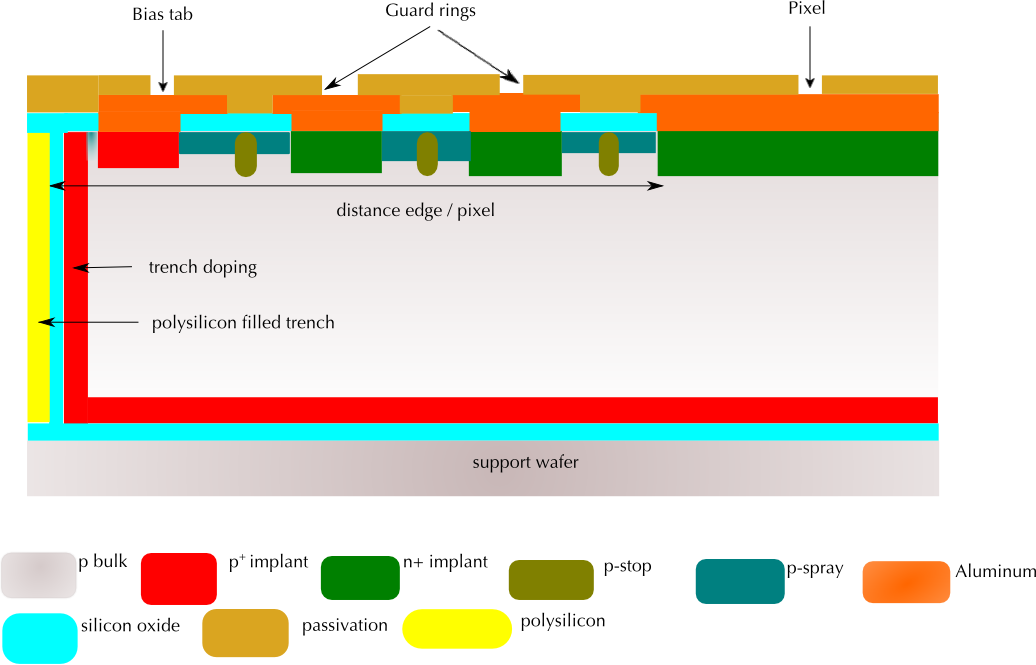}
\caption{\label{fig:pixel}Schematic section of the pixel sensor. The region close to the sensor's edge is portrayed, including the pixel closest to the edge, 
the edge region, including GRs (when present), the bias tab (present only on one edge of the device), the vertical doped trench, and the support wafer.}
\end{center}
\end{figure}

The trenches are then oxidized and filled with polysilicon. 
The remaining processing, arriving at the final device, whose cross-section is sketched in Figure~\ref{fig:pixel}, 
is quite standard, and includes the following steps: 
\begin{itemize}
\item contact opening
\item metal deposition and patterning
\item deposition of a passivation layer  (PECVD  oxide)  and patterning of the same in the
 pad and bump-bonding regions.
\end{itemize}

Since some sensors will be bump-bonded to  FE-I4~\cite{bib:fei4} read-out chips, 
it is necessary to select good sensors at the wafer level, by measuring their I-V characteristics.
 For this purpose, an additional layer of metal is deposited over the passivation and patterned into stripes, each of them shorting together a row of pixels, contacted through 
 the small passivation openings foreseen for the bump bonding.
This solution has already been adopted for the selection of good 3D FE-I4 sensors for the ATLAS IBL~\cite{bib:metal}.
After the automatic current-voltage  measurement 
 on each FE-I4 sensor, the metal will be removed by  wet etching, which does not affect the electrical characteristics  of the devices.

\section{The wafer layout}
\label{sec:prod}

\begin{figure}[!h]
\begin{center}
\includegraphics[height=0.29\textheight]{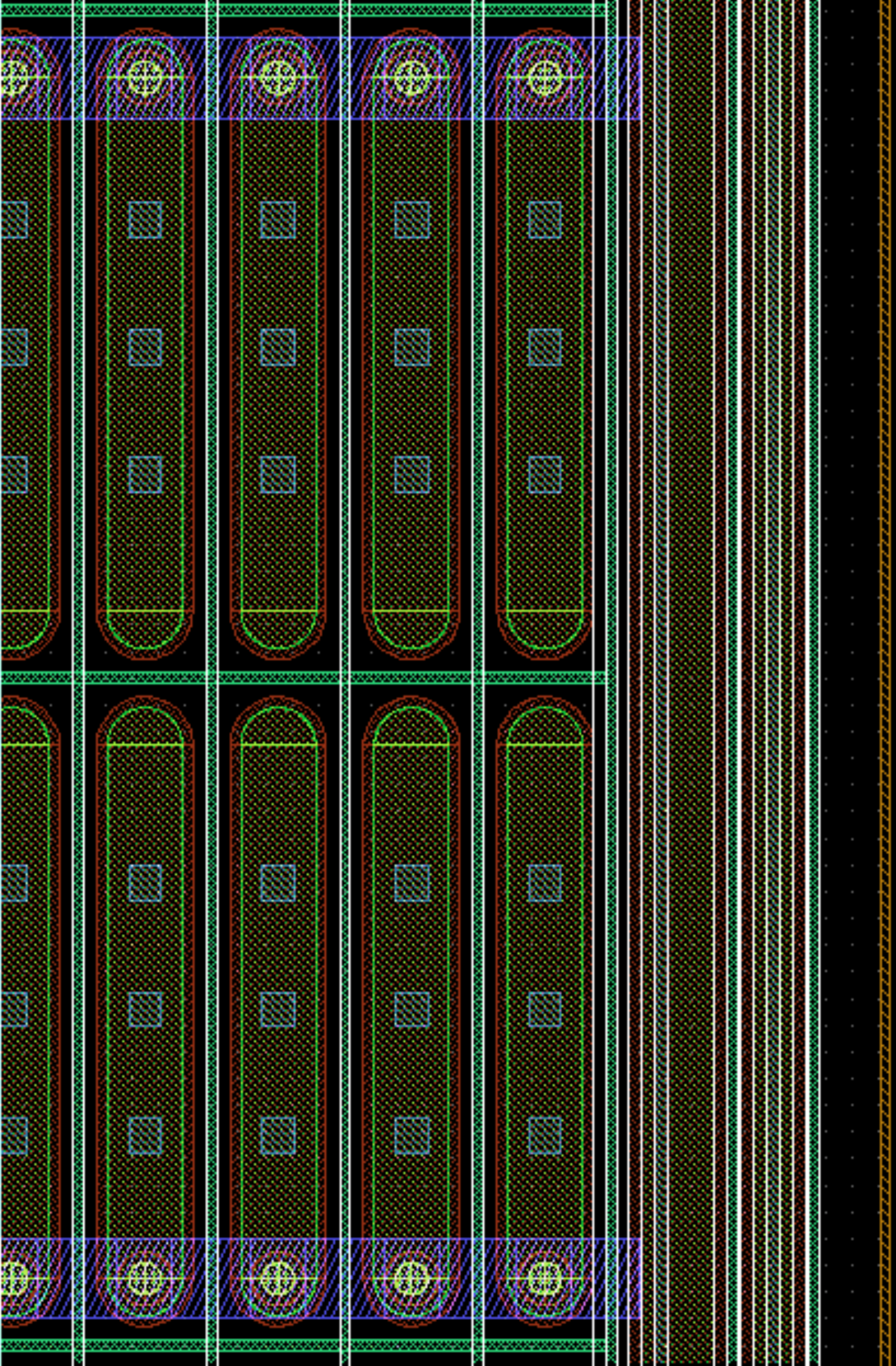}
\caption{\label{fig:gdsFEI4}Layout of a detail of an FE-I4 compatible pixel array.}
\end{center}
\end{figure}

FE-I4 compatible pixel sensors consist of an array of 336~$\times$~80 pixels, at a pitch  of 50~${\rm \mu}$m~$\times$~250~${\rm \mu}$m, 
for an overall sensitive area of 16.8~mm~$\times$~20.0~mm. Thus, on a 100~mm wafer, 
there is space for nine FE-I4 compatible pixel sensors; a detail of one of them is shown in Figure~\ref{fig:gdsFEI4}.

 The nine FE-I4 sensors differ in the pixel-to-trench distance (100, 200, 300, and 400~${\rm \mu}$m) and in the number of the guard rings (0, 1, 2, 3, 5, and 10)  
surrounding the pixel area (see Figure~\ref{fig:pixel}). The sensor with 3 GRs and a 200~$\mu$m pixel-to-trench distance features two different GR designs, and 
each of them is repeated twice. 
A list of the different FE-I4 sensor versions is reported in Table~\ref{tab:layout_split}.

\begin{savenotes}
\begin{table}[!ht]
\begin{center}
\begin{tabular}{ccc}

Multiplicity & Number of GRs & \begin{tabular}[x]{@{}c@{}}pixel-to-trench\\ distance (${\rm \mu m}$)\end{tabular}   \\
\hline
1 & 0 & 100 \\
1 & 1 & 100 \\ 
1 & 2 &100 \\
4 & 3 & 200 \\ 
1 & 5 & 300 \\ 
1 & 10 & 400  
\end{tabular}
\end{center}
\caption{\label{tab:layout_split}List of FEI4 sensors.  The number of the sensors (first column) is reported for each combination of number of GRs and pixel-to-trench distance. 
Two different designs are envisaged for the sensor with 3 GRs and 200~$\mu$m pixel-to-trench distance. See text for more details.}
\end{table} 
\end{savenotes}

A bias tab for substrate biasing (either by probing or by wire bonding), 
 located internally to the surface delimited by the trench, is placed at about 1.5~mm from the pixelated area on 
one of the sides (see also Figure~\ref{fig:pixel}).

The wafer layout also includes four sensors compatible with the FE-I3 read-out chip~\cite{bib:FEI3} (array of $160 \times 18$ pixels at a 
pitch of $50 {\rm \mu m} \times 400 {\rm \mu m}$, for an overall area of about 8~mm $\times$ 7.2 mm). The trench-pixel distance is 100~$ {\rm \mu m}$ for all of them, 
while they differ in the number of GRs (1 or 2).
In addition, four sensors compatible with the OmegaPIX readout chip~\cite{bib:omegapix} are  present (array of $96\times24$ pixels at a pitch of $35  {\rm \mu m} \times 200 
{\rm \mu m}$, 
for an overall area of about 3.4~mm  $\times$ 4.8 ~mm).

At the periphery of the wafer, there is room for a large number of test structures, {\it i.e.} square diodes and small arrays of FE-I4-like pixels, which differ in the number of 
GRs surrounding the active area and in the trench-to-pixel distance.  Several possible combinations  have been implemented, including all  those used for the FE-I4 sensors.
The aim of these structures is to  test the isolation and to measure the high-voltage behavior before and (possibly) after irradiation, in order to find the best sensor configuration
 to be  
bump-bonded to the read-out chip and to select the best combination of GR number and trench distance for a possible future production. 

To study the charge collection properties of the sensors,  particularly in the edge region, ``stripixel'' arrays have also been designed. 
They are  small 1-D arrays of about 2~mm long strips, which are an elongated version of the FE-I4 pixels  with the addition of a pad at one end, so that they 
can be wire bonded to a readout chip for strip detectors. Also in this case, 
there are several combinations of trench distances and number of GRs.

\section{TCAD simulation}
\label{sec:simu}

In order to explore and compare the properties of the design variations considered, numerical simulations were performed with TCAD tools from SILVACO~\cite{Silvaco}.
 2D structures analogous to the one sketched 
 in Figure~\ref{fig:pixel} have been simulated, varying parameters like the number of GRs and the pixel-to-trench distance. The break down (BD) 
 behaviour of the devices, the electrical field distribution and the charge collection efficiency (CCE) were studied, for simulated un-irradiated and irradiated 
 sensors, with a fluence  $\phi =  1 \times 10^{15} \rm{n_{eq}/cm^2}$.
 This is the expected fluence for the outer pixel layers  of the new tracker
 at the end of the HL-LHC phase.

 In the following,  details on dopant parameterization, on device physics models adopted and on the radiation damage parameterization will be presented, followed 
 by a selection of 
 results from simulations.
 
 \subsection{Doping parameters}
 \label{subsec:implants}
 
 Each of the doped regions (n$^+$ for the pixel and the GRs, p$^+$ for the backside, p-stop, p-spray, bias tab and the trench walls)  
 have been modeled  with simple functions, depending 
 on a set of parameters like the peak concentration, the reference concentration, {\it i.e.} the concentration value at a distance equal to the 
 ``rolloff''~\footnote{The doping concentration decreases from its peak value to its reference value over a distance equal to the rolloff} 
 from the peak position, and 
 the vertical ``rolloff'' distance. The  values used are summarized in Table~\ref{tab:implants}.

 \begin{table}[!ht]
\begin{center}
\begin{tabular}{lccccc}
Doped region & impurity & function & peak value (cm$^{-3}$)   & reference value (cm$^{-3}$)  & rolloff (${\rm \mu}$m) \\
\hline
	Pixel and GR  & D &   gaussian & $2\times10^{19}$ &   $10^{16}$ &  1.0 \\
	Back  &A &  gaussian & $2\times10^{19}$ &   $10^{16}$ &  1.0 \\
	Trench  &A &  erf & $2\times10^{19}$ &   $10^{12}$ &  2.0 \\
	Bias tab  &A &  gaussian & $2\times10^{19}$ &   $2\times10^{16}$ &  0.5 \\
	P-spray  & A &  gaussian & $5\times10^{16}$ &   $7\times10^{15}$ &  0.5  \\
	P-stop  & A &  gaussian & $5\times10^{17}$ &   $7\times10^{16}$ &  0.5  
\end{tabular}
\end{center}
\caption{\label{tab:implants}Implant parameters for simulated detectors; A (D) is for acceptor (donor) impurities.}
\end{table}

\subsection{Physics models and radiation damage parameterization}
\label{subsec:models}

SILVACO TCAD uses a complete set of physical models for semiconductor device simulation. Among them, models for concentration dependent mobility, 
field dependent mobility, bandgap narrowing, concentration dependent lifetime, trap-assisted and Auger recombination  were used.
Oxide fixed charge density  (with surface density $Q_{\rm ox}=10^{11}/{\rm cm^2}$ before irradiation, and $Q_{\rm ox}=3 \times 10^{12}/{\rm cm^2}$ after), 
  generation-recombination lifetimes and surface recombination velocity  have been set according to measured IV and CV characteristics of diodes from previous n-on-p 
CiS\footnote{Forschungsinstitut f\"ur Mikrosensorik und Photovoltaik GmbH}  productions. 

The defects at the edge have been modeled with a  1~$\mu$m wide region in which the generation-recombination lifetime was set to a very small value (10$^{-12}$~s; for 
comparison, before irradiation the corresponding value for the bulk is of $10^{-5}$~s). If the  trench doping were not effective, a large current would appear 
as soon as the electric field reaches the edge area.

To describe the radiation damage, an effective model based on three deep levels in the forbidden gap  was used~\cite{bib:Pennicard}. 
Each of these deep levels is defined as 
either donor (D) or acceptor (A), and is characterized by its energy (with respect to the closest energy band), its capture cross-sections for electrons
($\sigma_e$) and holes ($\sigma_h$) and its 
introduction rate $\eta$, which is the proportionality term between defect concentration and radiation fluence, expressed as 1~MeV n$_{\rm eq}$/cm$^2$. 
In Table~\ref{tab:Pennicard} 
 these properties are summarized.

\begin{table}[!ht]
\begin{center}
\begin{tabular}{ccccc}
  Type & Energy (eV) & $\sigma_e (\rm{cm}^2)$ & $\sigma_h (\rm{cm}^2)$ & $\eta  (\rm{cm}^{-1})$ \\ 
  \hline
  A &   $E_C $ -0.42 &  $9.5 \times10^{-15} $ &  $9.5 \times 10^{-14} $ & 1.613 \\
  A &   $E_C $ -0.46  &  $5.0 \times10^{-15} $ &  $5.0 \times 10^{-14} $ & 0.9 \\
  D &   $E_V $ +0.36  &  $3.23 \times10^{-13} $ &  $3.23 \times 10^{-14} $ & 0.9 
\end{tabular}
\end{center}
\caption{\label{tab:Pennicard}Relevant parameters for  acceptors (A) and donor (D) deep levels in the bandgap, describing the radiation damage.} 
\end{table} 

The 
deep level close to the centre of the bandgap is an highly effective generation center affecting the leakage current, 
while the other two contribute to the change of the effective doping concentration of the bulk and hence the depletion voltage.
  
Radiation-induced interface traps at the Si-SiO$_{\rm 2}$ interface are also included in the simulation, as described in~\cite{bib:InterfaceRD50}.

The model was validated by comparing simulation results to the change in depletion voltage and leakage current in irradiated n-on-p diodes from previous n-on-p CiS productions.   
  
\subsection{Simulation results}
\label{subsec:simresults}  

The structure in Figure~\ref{fig:pixel} was slightly modified in the simulations: the support wafer was not present and the backside p$^+$ implant 
was metallized.  
This was done in order to simulate a sensor ready for use. 

The sensors were simulated under reverse bias, applying a negative voltage to the back contact while keeping the pixel at ground potential;
  the 
bias tab was left floating. Different geometries were simulated, varying the number of GRs and the pixel-to-trench distance; see Table~\ref{tab:sim_devices} 
for the list of simulated geometries. If present, the GRs were left floating during the simulations.

\begin{savenotes}
\begin{table}[!ht]
\begin{center}
\begin{tabular}{cc}

\# of GRs & pixel-to-trench distance (${\rm \mu m}$) \\
\hline
 0 & 100 \\
1 & 100 \\ 
2 &100\\
 0 & 200\\ 
 1 & 200\\ 
 2 & 200\\ 
\end{tabular}
\end{center}
\caption{\label{tab:sim_devices}List of simulated sensor layouts.}
\end{table} 
\end{savenotes}

\subsubsection*{Current-voltage characteristic and break down voltage}

\begin{figure}[!htb]
\begin{center}
\includegraphics[width=0.85\textwidth]{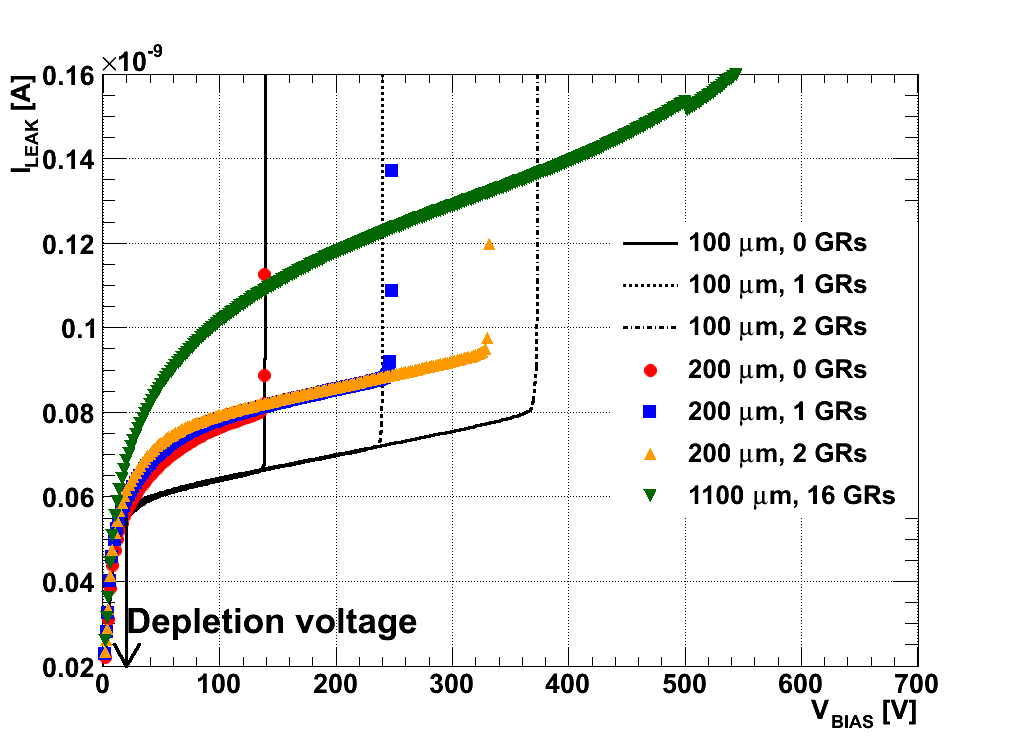}
\caption{\label{fig:BD}Simulated IV curves for the pixel closest to the edge, for several sensor designs before irradiation (see text for details). 
The simulated current has been scaled to reproduce the behaviour of a 50~${\rm \mu}$m wide pixel in the edge direction. 
The depletion voltage is indicated by the arrow.}
\end{center}
\end{figure}

Figure~\ref{fig:BD} shows the current-voltage curves of all the simulated designs, before irradiation. 
The depletion voltage has been estimated using AC analysis for simulations, determining the depletion voltage value from the fit to the $log(C)- log(V)$ curve; 
 the result was checked against the aforementioned measurements on 
n-on-p diodes from a former production. A sensor with a design compatible with the current ATLAS pixel modules was also simulated; it features a 
pixel-to-trench distance of 1.1~mm and 16 GRs. 
  
From Figure~\ref{fig:BD} it can be seen that before irradiation the BD voltage exceeds by at least 100~V the depletion voltage for all the designs we considered. 
The ATLAS-like sensor shows higher BD voltage with respect to those predicted for our edgeless detectors, 
but all sensors are largely over-depleted before BD. 
Increasing the pixel-to-trench distance yields a higher bulk-generated current, since the depleted volume can further extend laterally. Adding
more GRs 
helps greatly in increasing the value of BD voltage, extending the operability range of the sensors. The best performance is obtained from a device with 2 GRs 
and a $100 {\rm \mu}m$  pixel-to-trench distance. 
The BD voltage from simulations is  in agreement with other active edge productions; 
see for example~\cite{bib:VTT}~\footnote{The trenches are doped with a four-quadrant ion implantation
of boron ions.}

\begin{figure}[!htb]
\begin{center}
\includegraphics[width=0.85\textwidth]{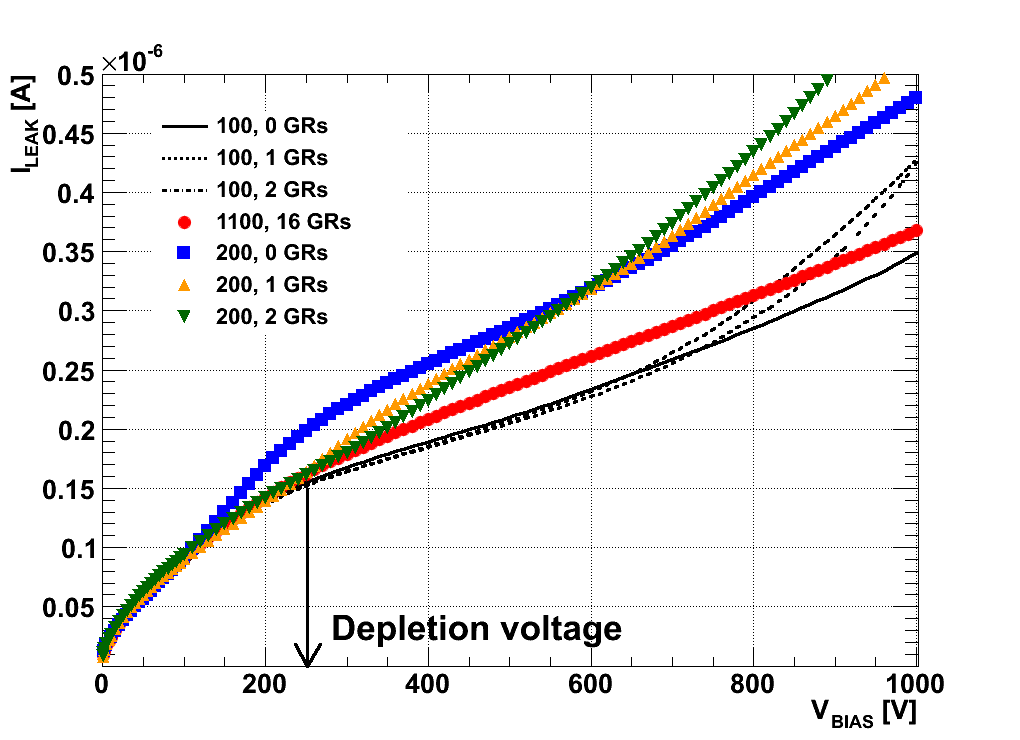}
\caption{\label{fig:BD1E15}Simulated IV curves for the pixel closest to the edge, for several sensor designs  after a simulated fluence of $10^{15} \rm{n_{eq}/cm^2}$ 
(see text for details). The simulated current has been scaled to reproduce the behaviour of a 50~${\rm \mu}$m wide pixel in the edge direction. 
The depletion voltage is indicated by the arrow.}
\end{center}
\end{figure}

As reported in the literature by different groups ({\it e.g.} \cite{bib:MPI}), after irradiation the BD voltage increases to much larger values. 
Our simulations of irradiated devices confirm this 
observation, as it can be seen in Figure~\ref{fig:BD1E15} where the same set of sensors of Figure~\ref{fig:BD} is now presented after a simulated fluence 
of $10^{15} \rm{n_{eq}/cm^2}$. No  BD is observed in any sensor up to 1000~V bias voltage.

\subsubsection*{Electric field distribution}

\begin{figure}[!htb]
\begin{center}
\includegraphics[width=0.89\textwidth]{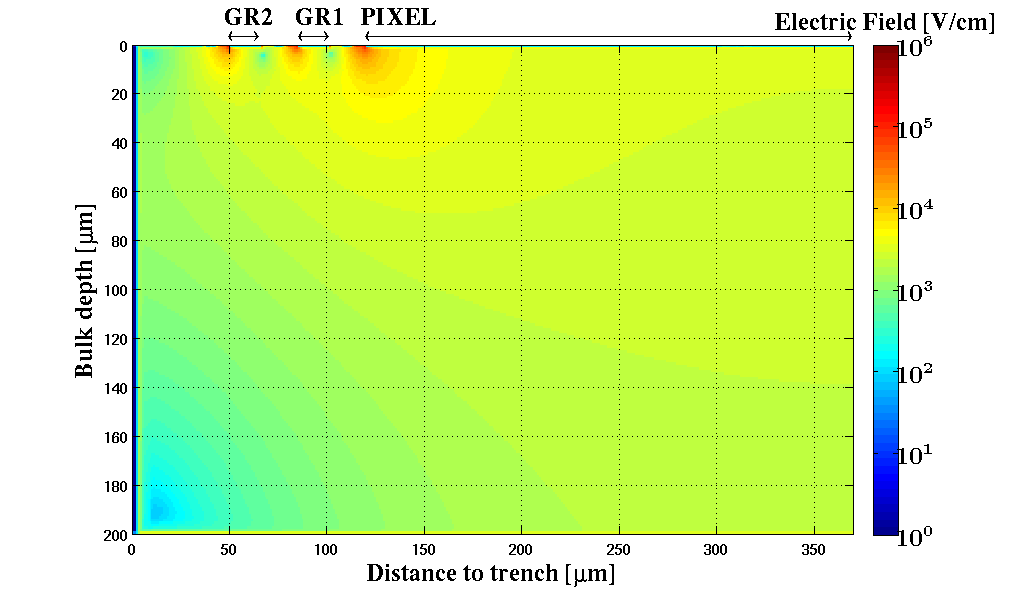}
\caption{\label{fig:Efield_fl0_50V}Electric field distribution for non-irradiated device at $\rm{V_{bias}=50\,V}$. The sensor has 2 GRs with a   100 $\rm{\mu m}$ distance between 
 trench and pixel.}
\end{center}
\end{figure}

In~Figure~\ref{fig:Efield_fl0_50V} the electric field distribution is shown for an un-irradiated detector with 2 GRs and $100{\rm \mu}$m  pixel-to-trench distance, for a 
bias voltage of 50~V.
It can be seen that the detector is fully depleted and the electric field is maximum in the region of the  p-stops, with a value of some units in $10^4 {\rm V/cm}$. 
It can also be observed that the electric field in the edge region is non negligible, hence signal should be collectable from there.

\begin{figure}[!ht]
\begin{center}
\includegraphics[width=0.89\textwidth]{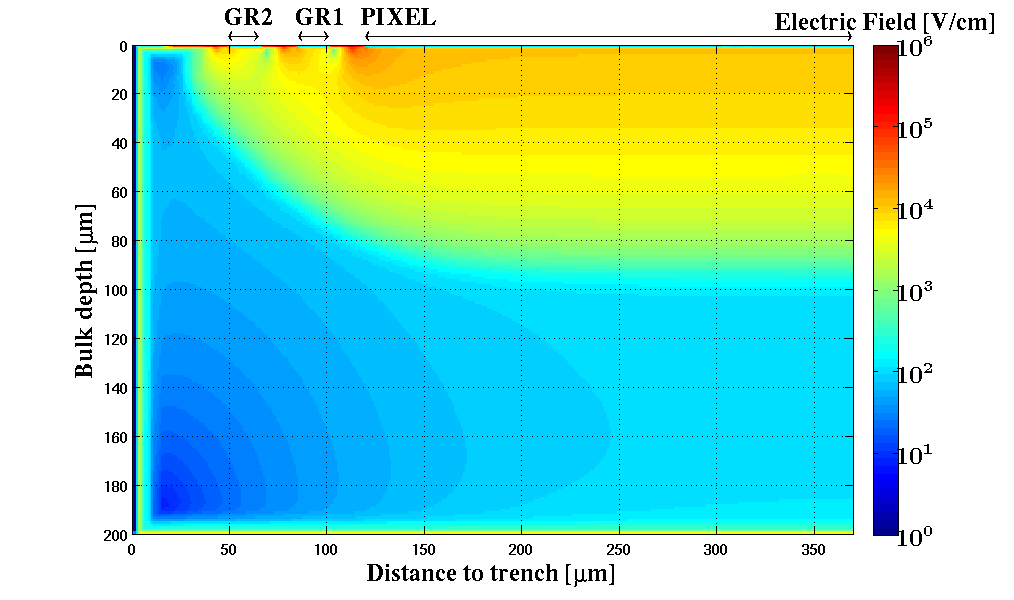}
\includegraphics[width=0.89\textwidth]{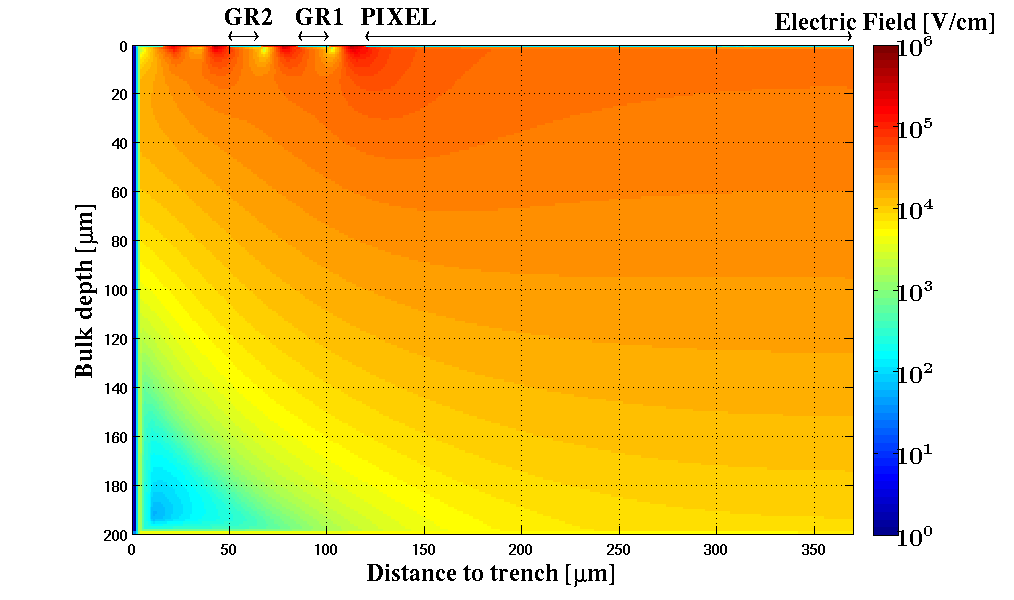}
\caption{\label{fig:Efield_fl1e15}Electric field distributions for a device irradiated with a fluence  $\phi=10^{15} \rm{n_{eq}/cm^2}$. Top: 
$\rm{V_{bias}=50\,V} $; bottom:  $\rm{V_{bias}=400\,V} $. The sensor has 2 GRs and a 100 $\rm{\mu m}$ distance between 
 edge and pixel.}
\end{center}
\end{figure}

In~Figure~\ref{fig:Efield_fl1e15} the electric field distribution is reported for a  sensor with 2 GRs 
and $100{\rm \mu}$m  pixel-to-trench distance, after a simulated fluence  $\phi = 10^{15} \rm{n_{eq}/cm^2}$.

In Figure~\ref{fig:Efield_fl1e15}, top, the detector is biased at 50~V; a large portion of the volume is non-depleted, {\it i.e.} the 
electric field is negligible. 
An interesting feature is an increase of the high electric field  close to the back implant and to the trench: this is a known effect called double peak (DP)~\cite{bib:DP}.
The fact that it is accounted for by our simulation supports the reliability of the simulation itself.

In~Figure~\ref{fig:Efield_fl1e15}, bottom, the electric field distribution is reported for the same irradiated detector at a bias voltage of 400~V, well beyond the depletion voltage 
($\sim 250$~V). The electric field extends all over the bulk, although a small undepleted region is still present at the edge, near the back-side; nonetheless, a large portion of the 
region between the pixel and the trench shows a sizable electric field, confirming the possibility of charge collection in the edge region, after irradiation too. 
As before irradiation, the electric field is maximum in the region of the  p-stops, but now its value is in the $10^5 {\rm V/cm}$ range.

\subsubsection*{Charge collection efficiency}

To study charge collection efficiency (CCE) after irradiation,  charge creation in irradiated sensors was simulated. The most interesting case is when the charge is 
released in the gap between the pixel and the trench, when no GRs are present. If a significant amount of charge can be collected after irradiation in that
 region, the edgeless concept would be verified to work.

Our sensor was illuminated from the front side with a simulated 1060~nm laser beam, setting its power in order to generate the same charge that would be 
released by a minimum ionizing particle (MIP) traversing 200~$\mu$m of silicon ($\sim 2.6$~fC). The laser beam was originating above the front side of the 
detector, with a 2~$\mu$m wide gaussian beamspot. 
 The duty cycle of the laser was  50~ns, with the power ramping up in 1~ns, remaining constant for 10~ns and ramping down in the next nanosecond.

The CCE was studied as a function of the bias voltage for the detector with no GRs and a 100~$\mu$m  trench-to-pixel distance. 
Two incidence points of the laser beam have been considered: one within the pixel and the other in the edge region, at 50~$\mu$m distance from the pixel. 
In the following they will be identified as ``Pixel'' and ``Edge'', respectively.

Based on the properties of the laser beam and of the target material, the simulation program determined the charge of carriers photogenerated inside the device by one pulse. 
The charge collected by the pixel was defined as the integral over the laser duty cycle of the current flowing through the pixel, 
once the stable leakage current had been subtracted. Finally, the CCE was obtained by dividing this collected charge by the total photogenerated charge.

\begin{figure}[!ht]
\begin{center}
\includegraphics[width=0.89\textwidth]{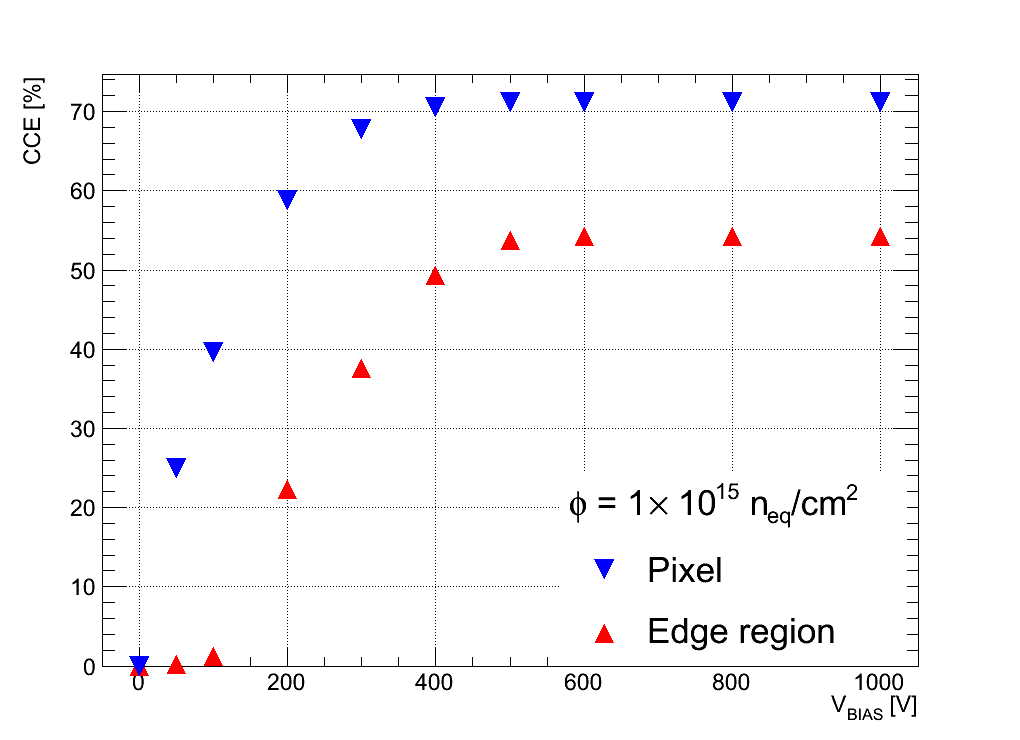}
\caption{\label{fig:CCE_comp_100_0gr}Charge collection efficiency as a function of bias voltage for an irradiated device at a fluence  $\phi= 10^{15} \rm{n_{eq}/cm^2}$ . 
The laser is entering 
the detector either in the pixel region (``Pixel'') or in the un-instrumented region (``Edge region'').
 The sensor has no GRs, and a 100 $\rm{\mu m}$ distance between edge and pixel.}
\end{center}
\end{figure}

In Figure~\ref{fig:CCE_comp_100_0gr} the CCE is presented as a function of the bias voltage for the  simulated fluence for the two incidence points of the laser beam. 

At a fluence  $\phi = 10^{15} \rm{n_{eq}/cm^2}$  more than 50~\% of the signal  is collected in the ``Edge'' region at a bias voltage of 500~V; 
as a comparison, 70~\% of the  signal is retained in the ``Pixel'' region.  The expected collected charge in the ``Edge'' region 
is then of $\sim$8~ke~\footnote{the MPV for the charge created by a MIP in 200~$\rm{\mu m}$ is 16~ke}, 
which corresponds to  
a signal large enough to trigger the FEI4 readout chip.
Both in the ``Pixel'' and in the ``Edge'' region the effect of trapping can be observed: the collected charge reaches a {\it plateau} at high voltage, but there the CCE is not of 100~\%. 
No charge is collected from the ``Edge'' region below 100~V: indeed at 100~V bias the electric field is negligible in that region when there are no GRs.  
It can be seen that while the maximum CCE for a charge created in the pixel region is reached at a bias voltage above $\sim 400$~V,  in the 
``Edge'' region a bias voltage of 600~V is needed: this is consistent with the depletion zone extending laterally.


Calculations based on trapping time experimental data~\cite{bib:Trapping} for our sensor thickness and  our target fluence produce CCE estimations
  in agreement with our simulations.

\section{Conclusions and outlook}

In view of the Large Hadron Collider High Luminosity (HL-LHC) phase, an upgrade of the ATLAS Inner Detector is envisaged. New pixel sensors will have to work in  
an unprecedentedly harsh environment; moreover,  material budget restrictions and  
 tight mechanical constraints demand for a reduction of the inactive region at the edge of sensors. 
 
 FBK Trento and LPNHE Paris are developing new planar n-on-p pixel sensors for the ATLAS detector upgrade, characterized by a reduced inactive region at 
 the edge thanks to a vertical doped lateral surface at the device boundary, the ``active edge'' technology. 
 Simulation studies show the effectiveness of this technique in reducing the dead area while retaining a large margin of operability in terms of bias voltage, even after 
 simulated fluences comparable to those expected at the end of the HL-LHC phase for the external layers; they also show that  after irradiation 
more than 50~\% of the signal is retained, 
even in the  ``Edge'' region for the sensor with 100~$\rm{\mu m}$  pixel-to-trench distance.
 
Next steps will be a full electrical characterization of these ``active edge'' devices, followed by functional tests of the pixel sensors  with radioactive sources 
 and eventually in a beam test, after having bump bonded a number of  pixel sensors to the FE-I4 read out chips.

\section*{Acknowledgements} 
The authors would like to express their gratitude to E.~Fretwurst for his useful discussions. 
This work was supported in part by the Autonomous Province of Trento, Project MEMS2, and
in part by the Italian National Institute for Nuclear Physics (INFN).












\end{document}